\documentclass[proceedings]{JHEP} 

\conference{Corfu Summer Institute on Elementary Particle Physics, 1998}

\title{Top Mixing}

\author{F. del Aguila and J. A. Aguilar-Saavedra \\
        Departamento de F\'{\i}sica Te\'{o}rica y del Cosmos,
        Universidad de Granada, 18071 Granada, Spain\\
        E-mail: \email{faguila@ugr.es} and \email{aguilarj@ugr.es}}

\abstract{We discuss the existing limits on top flavour-changing neutral
couplings in models with new vector-like quarks. Large hadron and $e^+ e^-$
colliders can improve these bounds by more than one order of magnitude.}

\begin{document} 

\section{Introduction}
The top quark is the heaviest fermion discovered up to now.
Its mass $m_t = 173.8 \pm 5.2$
GeV \cite{papiro2} is $3.5 \times 10^5$ times the electron mass and
makes more demanding
the question of why the fermion mass hierarchy is so large. Not to say
if we compare $m_t$ with the light neutrino masses. At present there is no
compelling theoretical reason for such spreading. The Standard Model (SM) simply
accommodates fermion masses and mixings. One is tempted, however, to relate the
large value of the top mass, which is not very different from the electroweak
symmetry breaking scale, with the mass generation mechanism. This and any
other new physics are then expected to show up at the top first.

In contrast, top couplings are constrained rather poorly by present experimental
data, especially the flavour-changing neutral (FCN) ones. In the SM there are no
tree level FCN couplings and the effective ones induced at one loop are
negligible. However, the top quark can have large tree level FCN couplings in
models with extra vector-like quarks \cite{papiro10,papiro10b,papiro11}.
(These are fermions whose left-handed (LH)
and right-handed (RH) parts have the same transformation properties
under $\mathrm{SU}(2)_L$. They are present in many extensions of the SM and are
the only fermions whose mass can be banished to high energy.)
Hence, it is theoretically important to constrain them experimentally.
In Sections 2 and 3 we discuss the existing limits on these couplings in these
simple and well-defined SM extensions \cite{papiro5}.
In Section 4 we comment on the bounds
which can be obtained at large hadron and $e^+ e^-$ colliders. Existing limits
will be eventually improved by at least one order of magnitude
\cite{papiro6,papiro6b,papiro6c,papiro7,papiro8}.

\section{Model-independent limits}
Direct limits on top FCN couplings can be derived from top decays. At present
the most stringent bounds result from the non-observation of the decays $t \to
Zq$, $t \to \gamma q$, $t \to gq$ at the Fermilab Tevatron.
The neutral trilinear interactions between the top and a light quark $q$
are given by the Lagrangian
\begin{eqnarray}
\mathcal{L}_\mathrm{FCNC} & = 
& \bar t \left[ \frac{g}{2 c_W}  \gamma_\mu (X_{tq}^L P_L +
 X_{tq}^R P_R)Z^\mu \right. \nonumber \\
& & + \frac{g}{2 c_W} (\kappa_{tq}^{Z}-i \tilde \kappa_{tq}^{Z})
\frac{i \sigma_{\mu \nu} q^\nu}{m_t} Z^\mu  \nonumber \\
& &  + e (\kappa_{tq}^{\gamma} - i \tilde \kappa_{tq}^{\gamma})
\frac{i \sigma_{\mu \nu} q^\nu}{m_t} A^\mu \nonumber \\
& & + \left. g_s (\kappa_{tq}^{g} -i \tilde \kappa_{tq}^{g})
\frac{i \sigma_{\mu \nu} q^\nu}{m_t} T^a
G^{a\mu} \right] q \nonumber \\
& & +{\mathrm h.c.}\,, \label{ec:1}
\end{eqnarray}
with $P_{R,L}=(1 \pm \gamma_5)/2$ and $T^a$ the Gell-Mann matrices
satisfying ${\mathrm Tr}\, (T^a T^b) = \delta^{ab}/2$.
Tevatron bounds \cite{papiro3,papiro4}
\begin{eqnarray}
\mathrm{Br}(t \to Zq)       & \leq & 33\% \nonumber \\
\mathrm{Br}(t \to \gamma q) & \leq & 3.2\% \nonumber \\
\mathrm{Br}(t \to gq)       & \leq & 15\%
\end{eqnarray}
translate into $|X_{tq}^L|,|X_{tq}^R| \leq 0.84$,
$|\kappa_{tq}^{Z}|,|{\tilde \kappa}_{tq}^{Z}| \leq 0.78$,
$|\kappa_{tq}^{\gamma}|,|{\tilde \kappa}_{tq}^{\gamma}| \leq 0.26$, 
$|\kappa_{tq}^{g}|,|{\tilde \kappa}_{tq}^{g}| \leq 0.15$.

The effective couplings proportional to
$\sigma_{\mu \nu}$ in Eq.~\ref{ec:1} are absent at tree level in renormalizable
theories. Hence they are suppressed by one-loop factors $\sim \alpha/\pi$, and
also by the GIM mechanism if there are no tree-level $(\gamma_\mu )$ FCN
couplings. Thus we will neglect them in the following. However, SM extensions
with vector-like quarks
allow for large (tree-level) $\gamma_\mu$ FCN couplings. This is so because the
addition of vector-like quarks has little effect on radiative corrections. In
particular, the new quarks decouple when their masses are taken to infinity.
Although FCN couplings are constrained by precise electroweak data,
indirect limits are at the mercy of possible cancellations between the top and
other new particle contributions (see Ref. \cite{papiro6} for an effective
Lagrangian approach). It may be then convenient to derive these constraints in
specific models. In the next Section we discuss these limits for SM extensions
with vector-like fermions.

\section{Limits in extended models with vector-like quarks}
Let us consider a general SM extension with $N$ standard families, $n$
vector-like doublets and $n_u$ up and $n_d$ down vector-like singlets.
(In practice, our limits will be saturated in the minimal
extensions with $N=3$ and one extra doublet or singlet.) The quark content of
this model is summarized in Table~\ref{tab:1}. The weak eigenstates are then 
$q_{L,R}^0=(q_{L,R}^{(d)},q_{L,R}^{(s)})$, with $q_{L,R}^{(d)}$
doublets and $q_{L,R}^{(s)}$ singlets, and $q=u,d$.

\begin{table}[h]
\begin{center}
\begin{tabular}{lc}
LH doublets & $N+n$ \\
RH doublets & $n$ \\
up LH singlets & $n_u$ \\
up RH singlets & $N+n_u$ \\
down LH singlets & $n_d$ \\
down RH singlets & $N+n_d$
\end{tabular}
\end{center}
\caption{Quark content of the model \label{tab:1}}
\end{table}

Note that the total number of up-type quarks $\mathcal{N}_u = N+n+n_u$
 and down-type
quarks $\mathcal{N}_d = N+n+n_d$ do not need to be equal in general. As in
the SM, only isodoublets have $\mathrm{SU}(2)_L$ couplings,
then the gauge neutral current
Lagrangian in the weak eigenstate basis can be written as
\begin{eqnarray}
\mathcal{L}_{Z} & = & - \frac{g}{2 c_W} \left( \bar u_{L}^{(d)}
\gamma^\mu u_{L}^{(d)} + \bar u_{R}^{(d)} \gamma^\mu u_{R}^{(d)} \right.
\nonumber \\
& &  - \bar d_{L}^{(d)} \gamma^\mu d_{L}^{(d)} - \bar d_{R}^{(d)} 
\gamma^\mu d_{R}^{(d)} \nonumber \\
& & \left. - 2 s_W^2 J_\mathrm{EM}^\mu \right) Z_\mu \,.
\label{ec:3}
\end{eqnarray}
With this notation the only apparent difference with the SM Lagrangian
is in the new RH currents. However, when
we express this Lagrangian in the mass eigenstate basis the particular features
of this model manifest. In this basis,
\begin{eqnarray}
\mathcal{L}_{Z} & = & - \frac{g}{2 c_W} \left(
\bar u_{L} X^{uL} \gamma^\mu u_{L} +
\bar u_{R} X^{uR} \gamma^\mu u_{R} \right. 
\nonumber \\
& &  - \bar d_{L} X^{dL} \gamma^\mu d_{L} - 
\bar d_{R} X^{dR} \gamma^\mu d_{R} \nonumber \\
& & \left. - 2 s_W^2 J_\mathrm{EM}^\mu \right) Z_\mu \,,
\label{ec:4}
\end{eqnarray}
where $u=(u,c,t,T,\dots)$ and $d=(d,s,b,B,\dots)$ are ${\mathcal N}_u$ and
${\mathcal N}_d$ dimensional vectors, respectively. The $X$'s are matrices in
flavour space. Apart from the new RH terms, the most important difference with
the SM Lagrangian is that now the LH currents are not flavour diagonal, {\em i.
e. } the $X$'s are nondiagonal. To show this, let us write the unitary
transformation between the mass and weak interaction eigenstates, 
$q_{L}^0=\mathcal{U}^{qL} q_{L}$,
$q_{R}^0=\mathcal{U}^{qR} q_{R}$, with $\mathcal{U}^{qL}$ and
$\mathcal{U}^{qR}$ $\mathcal{N}_q \times \mathcal{N}_q$ unitary matrices.
Then,
\begin{eqnarray}
X_{\alpha \beta}^{uL} & = &
(\mathcal{U}^{uL}_{i\alpha})^* \mathcal{U}^{uL}_{i\beta} \,, \nonumber \\
X_{\alpha \beta}^{uR} & = &
(\mathcal{U}^{uR}_{j\alpha})^* \mathcal{U}^{uR}_{j\beta} \,, \nonumber \\
X_{\sigma \tau}^{dL} & = &
(\mathcal{U}^{dL}_{k\sigma})^* \mathcal{U}^{dL}_{k\tau} \,, \nonumber \\
X_{\sigma \tau}^{dR} & = &
(\mathcal{U}^{dR}_{l\sigma})^* \mathcal{U}^{dR}_{l\tau} \,,
\label{ec:5}
\end{eqnarray}
where $(i,k)$ and $(j,l)$ sum over the left- and right-handed doublets,
respectively, $\alpha,\beta=u,c,t,T,\dots$ and $\sigma,\tau=d,s,b,B,\dots$ In
the SM, $\mathcal{U}^{qL}$, $\mathcal{U}^{qR}$ are $3 \times 3$ matrices and
$i,k$ run from 1 to 3, hence $X_{qq'}^{uL,dL}=\delta_{qq'}$ by
unitarity for any mass eigenstates $q$, $q'$.
This is the well-known GIM mechanism. Besides, $X_{qq'}^{uR,dR}
= 0$ for there are no RH doublets. In general, with $n>0$ and
$N+n<\mathcal{N}_u,\mathcal{N}_d$ the $X$'s are products of
submatrices with less rows than columns and
thus they are nondiagonal in general.

At this point one would have to wonder about the constraints on the $X$'s from
precise electroweak data, and in particular from rare processes. However, 
from the apparently trivial Eqs.~\ref{ec:5} we will find a set of inequalities
which imply limits on the top couplings strong
enough to fulfil all experimental data. To be
definite, we work with $\mathcal{U}^{uL}$, and to simplify the notation, we
write
\begin{equation}
\mathcal{U}^{uL} = \left(
\begin{array}{ccccc}
a_u & a_c & a_t & \cdots & a_{\mathcal{N}_u} \\
b_u & b_c & b_t & \cdots & b_{\mathcal{N}_u} 
\end{array}
\right) 
\begin{array}{l}
\} N+n \\
\} n_u
\end{array}\,,
\label{ec:6}
\end{equation}
with $a_\alpha$, $b_\alpha$ column vectors
(remember that $N+n$ is the number of up LH doublets and $n_u$ the number of
up LH singlets). With this notation, orthogonality between columns of
$\mathcal{U}^{uL}$ is written as $a_\alpha\cdot a_\beta +
b_\alpha \cdot b_\beta = \delta_{\alpha \beta}$, and $X^{uL}$ can be written as 
$X_{\alpha \beta}^{uL} = a_\alpha \cdot a_\beta$ with the complex scalar
product `$\cdot$' (in particular this
shows that $|X_{\alpha \beta}^{uL}| \leq 1$ and
 $X_{\alpha \alpha}^{uL} \geq 0$). It is
then easy to apply the Schwarz inequality to obtain
\begin{eqnarray}
|X_{\alpha \beta}^{uL}|^2 & = & |a_\alpha \cdot a_\beta| \leq |a_\alpha|^2
|a_\beta|^2 \nonumber \\
& = & X_{\alpha \alpha}^{uL} X_{\beta \beta}^{uL}\,,
\label{ec:7}
\end{eqnarray}
and for $\alpha \neq \beta$,
\begin{eqnarray}
|X_{\alpha \beta}^{uL}|^2 & = & |a_\alpha \cdot a_\beta| = |b_\alpha \cdot
b_\beta| \leq |b_\alpha|^2 |b_\beta|^2 \nonumber \\
& = & (1-|a_\alpha|^2) (1-|a_\beta|^2) \nonumber \\
& = & (1-X_{\alpha \alpha}^{uL}) (1-X_{\beta \beta}^{uL}) \,.
\label{ec:8}
\end{eqnarray}
The same can be done for down quarks and RH couplings,
so we drop superscripts and
write in complete generality
\begin{eqnarray}
|X_{qq'}|^2 & \leq & (1-X_{qq}) (1-X_{q'q'}) \,,
\nonumber \\
|X_{qq'}|^2 & \leq & X_{qq} X_{q'q'} \,.
\label{ec:9}
\end{eqnarray}
It is important to remark that these inequalities are valid for any number of
generations $N$, doublets $n$ and singlets $n_u$, $n_d$.
In particular, if $X_{qq}=0,1$, Eqs.~\ref{ec:9} imply $X_{qq'}=0$ independently
of $X_{q'q'}$. In this way, Eqs.~\ref{ec:9} can be seen as the generalization of
the GIM mechanism for models with vector-like quarks.
In the case of the top quark, these equations allow to derive bounds on
$X_{tu}$, $X_{tc}$, despite our ignorance on $X_{tt}$. 

To use the inequalities in Eqs.~\ref{ec:9} we have to extract the values of the
diagonal elements $X_{qq}$ from experimental data. From atomic parity violation
\cite{papiro2,papiro20} and the SLAC polarized electron experiment
\cite{papiro21} the diagonal $X$ elements of the $u$ and $d$ quarks in
Table~\ref{tab:2} can be extracted, whereas the measurement of $R_b$, $R_c$,
$A_\mathrm{FB}^{0,b}$, $A_\mathrm{FB}^{0,c}$ at the CERN $e^+ e^-$ collider
LEP and SLC \cite{papiro22} 
provides the diagonal $X$
elements of the $b$ and $c$ quarks (see Ref. \cite{papiro5} for details).
We observe that the values of $X_{uu}^R$, $X_{dd}^L$, $X_{cc}^R$ and $X_{bb}^R$
in Table~\ref{tab:2} are unphysical. This is worst for
$X_{bb}^R$, which is 2$\sigma$ away from the physical region $[0,1]$, a direct
consequence of the 2$\sigma$ discrepancy between the measured and the SM values
of $A_{FB}^{0,b}$. It is then necessary a careful application of the
inequalities. We define the $90\%$ C. L.
upper limit on $X_{qq'}$ as the value $x$ such that the probability of finding
$X_{qq'} \leq x$ {\em within} the physical region is 0.9. With this definition
and a Monte Carlo generator for the Gaussian distributions of $R_b$, $R_c$,
$A_{FB}^{0,b}$, $A_{FB}^{0,c}$ (correlated) and $X_{uu}^{L,R}$, $X_{dd}^{L,R}$
(assuming no correlation) we obtain the bounds in Table \ref{tab:3}, where we
also quote existing limits in the literature \cite{papiro11,papiro3,papiro12}.

\begin{table}[h]
\begin{center}
\begin{tabular}{l}
$X_{uu}^L = 0.965 \pm 0.032$ \\ $X_{uu}^R = -0.049 \pm 0.026$ \\
$X_{dd}^L = 1.035 \pm 0.022$ \\ $X_{dd}^R = 0.209^{+0.096}_{-0.154}$ \\
$X_{cc}^L = 0.998 \pm 0.013$ \\ $X_{cc}^R = -0.013 \pm 0.019$ \\
$X_{bb}^L = 0.996 \pm 0.005$ \\ $X_{bb}^R = -0.039 \pm 0.018$
\end{tabular}
\end{center}
\caption{Diagonal elements in Eq.~\ref{ec:4} \label{tab:2} }
\end{table}

\TABLE{
\begin{tabular}{cccl}
Coupling & $X^L$ & $X^R$ & Source \\ \hline 
$uc$ & $1.2 \times 10^{-3}$ & $1.2 \times 10^{-3}$ &
$\delta m_D$ \\ 
 & 0.033 & 0.019 & Inequalities \\ \hline
$ut$ & 0.84 & 0.84 & $t \rightarrow uZ$ \\
 & 0.28 & 0.14 & Inequalities \\ \hline
$ct$ & 0.84 & 0.84 & $t \rightarrow cZ$ \\
 & 0.14 & 0.16 & Inequalities \\ \hline
$ds$ & $4.1 \times 10^{-5}$ & $4.1 \times 10^{-5}$ & 
$K^+ \rightarrow \pi^+ \nu \bar \nu$ \\
 & 0.14 & 0.62 & Inequalities \\ \hline
$db$ & $1.1 \times 10^{-3}$ & $1.1 \times 10^{-3}$ &
$\delta m_B$ \\
 & 0.0081 &  0.062 & Inequalities \\ \hline
$sb$ & $1.9 \times 10^{-3}$ & $1.9 \times 10^{-3}$ &
$B^0 \rightarrow \mu^+ \mu^- X$ \\
 & 0.076 & 0.12 & Inequalities
\end{tabular}
\caption{Present bounds on nondiagonal elements in Eq.~\ref{ec:4} \label{tab:3}}
}

It is worth to emphasize that with Eqs.~\ref{ec:9} and precise electroweak
data we can set limits on top couplings that ({\em i\/}) are much stronger than
present direct limits, and ({\em ii\/}) can be saturated in minimal SM
extensions with one doublet or singlet and still fulfil all present experimental
constraints \cite{papiro5}.

\section{Future limits}
Improvements on the determination of the top FCN couplings should come
from large hadron and $e^+ e^-$ colliders. At these top factories the top FCN
vertices manifest in top decays and in single top production. For $\sigma_{\mu
\nu}$ vertices and large center of mass
energies single top production is a more efficient
way to look for FCN couplings because the extra $q^\nu$ factor in the vertex
increases the large transverse momentum distributions improving the signal to
background ratios, whereas in top decays the energy available is typically
smaller and the characteristic $q^\nu$ factor makes no appreciable difference.
For definiteness, we restrict ourselves to renormalizable $Ztq$ couplings, which
are the relevant ones in extended models with vector-like quarks. (The other
couplings can be discussed in an analogous way.) $e^+ e^-$ machines are a
cleaner environment but less efficient top factories. LEP2 can improve present
Tevatron limits at most by a factor of 3 \cite{papiro22b}
and Tevatron Run II with a luminosity of 2 fb$^{-1}$
will improve LEP2 bounds by a factor of 2 \cite{papiro8}.
Finally, LHC with a luminosity of 100 fb$^{-1}$ will improve the former bounds
by a factor of 20 \cite{papiro23}. Linear colliders will have
to be consistent with the LHC limits but are not expected to improve them.
However, if a sizeable coupling is found, linear colliders will be able to
disentangle better the P and CP structure of the vertex.

\acknowledgments
We thank Ll. Ametller, G. C. Branco and R. Miquel for previous collaboration on
this subject. F. A. thanks the
organizers of the Summer Institute for their hospitality. 
This work was partially supported by CICYT under contract AEN96--1672 and by the
Junta de Andaluc\'{\i}a, FQM101.

\end{document}